# Magnetic Structure and Ordering of Multiferroic Hexagonal LuFeO$_3$

Steven M. Disseler[1], Julie A. Borchers[1], Charles M. Brooks[2], Julia A. Mundy[3], Jarrett A. Moyer[4], D. A. Hillsberry[5], E. L. Thies[5], D. A. Tenne[5], John Heron[2], James D. Clarkson[6], Gregory M. Stiehl[7], Peter Schiffer[4], David A. Muller[3,8], Darrell G. Schlom[2,8], and William D. Ratcliff[1*]

[1]NIST Center for Neutron Research, National Institute of Standards and Technology, Gaithersburg, Maryland, 20899, USA

[2]Department of Materials Science and Engineering, Cornell University, Ithaca, New York 14853, USA

[3]School of Applied and Engineering Physics, Cornell University, Ithaca, New York 14853, USA

[4]Department of Physics and Frederick Seitz Materials Research Laboratory, University of Illinois at Urbana-Champaign, Urbana, Illinois 61801, USA

[5]Department of Physics, Boise State University, Boise ID 83725

[6]Department of Materials Science and Engineering, University of California, Berkeley, California 94720, USA

[7]Department of Physics, Cornell University, Ithaca, New York 14853, USA

[8]Kavli Institute at Cornell for Nanoscale Science, Ithaca, New York 14853, USA

We report on the magnetic structure and ordering of hexagonal LuFeO$_3$ films grown by molecular-beam epitaxy (MBE) on YSZ (111) and Al$_2$O$_3$ (0001) substrates. Using a set of complementary probes including neutron diffraction, we find that the system magnetically orders into a ferromagnetically-canted antiferromagnetic state via a single transition between 138-155 K, while a paraelectric to ferroelectric transition occurs above 1000 K. The symmetry of the magnetic structure in the ferroelectric state implies that this material is a strong candidate for linear magnetoelectric coupling and control of the ferromagnetic moment directly by an electric field.

Multiferroic materials display both ferroelectric and magnetic order and have been the subject of intense investigation both from fundamental and applied perspectives [1,2]. For example, if both order parameters are coupled, these materials would enable new devices ranging from magnetic field sensors to magnetic random access memory. Unfortunately, single-phase multiferroics are extraordinarily rare; thus far only four room-temperature single-phase multiferroic have been reported: $BiFeO_3$ [3], $BiCoO_3$ [4], $ScFeO_3$ with the corundum structure [5], and most recently hexagonal $LuFeO_3$ ($h$-$LuFeO_3$) [6]. The latter compound, $h$-$LuFeO_3$ has been synthesized and stabilized in thin film form and was found to be ferroelectric and isostructural with $YMnO_3$ (Fig. 1(a)) [7, 8].

$YMnO_3$ and other hexagonal manganites, $REMnO_3$ (RE=Lu, Y, Ho), have been known for some time to exhibit multiferroic properties. The ferroelectric transition in these materials is the result of a structural transition from the non-polar $P6_3/mmc$ to the polar $P6_3cm$ space group well above room temperature ($T_C \sim 900$ K in $YMnO_3$) [9]. Magnetic order sets in, however, at much lower temperatures ($T_N \sim 80$ K in $YMnO_3$) [10], which has rendered these materials unsuitable for multiferroic device applications. Even in the magnetically ordered state at cryogenic temperatures, the coupling between the ferroelectric and magnetic order parameters is weak [11]. Replacing Mn with Fe in this system has been proposed as one way to increase both the magnetic transition temperature as well as the coupling between the two order parameters [12], and has been the subject of increased interest as of late [6, 13]. For example, theoretical calculations using first-principles suggest that a weak ferromagnetic moment the $c$-axis may be deterministically switchable by 180° with an electric field [12].

Reports of antiferromagnetic order at room temperature in $h$-LuFeO$_3$ [6] coupled with theoretical predictions of a weak canted moment along the $c$-axis, which is observed at temperatures below 147 K [13], suggest that this material could be of interest for device applications. In this Letter, we determine the intrinsic magnetic structure of $h$-LuFeO$_3$ epitaxial films through magnetometry and neutron diffraction measurements. We find that antiferromagnetic order is evident as previously reported [6] but occurs only below 155 K for $h$-LuFeO$_3$ on several substrates (i.e., Al$_2$O$_3$ and cubic zirconia). Further, its onset occurs simultaneously with the onset of a weak ferromagnetic canting of the moments. From Raman scattering we demonstrate that $h$-LuFeO$_3$ is ferroelectric at room temperature, with a paraelectric-to-ferroelectric transition temperature $T_C$ = 1020 K ± 50 K. Solving the magnetic structure we confirm that the films magnetically order in the ferroelectric state in a manner that is completely consistent with theoretical predictions for the magnetoelectric effect such that electric field-induced reversal of the ferromagnetic moment direction should be achievable in this material.

We used oxide molecular-beam epitaxy (MBE) to grow four high quality, single-crystalline samples of $h$-LuFeO$_3$ film on 10 mm × 10 mm substrates of either (111)-oriented yttria-stabilized cubic zirconia (YSZ) or (0001) Al$_2$O$_3$ [13]. Films of 200 nm and 250 nm thickness were grown on each type of substrate. For simplicity, the following scheme will be used to describe each sample in the remainder of this work: YSZ-200 nm (Sample 1), YSZ-250 nm (Sample 2), Al$_2$O$_3$-200 nm (Sample 3), and Al$_2$O$_3$-250 nm (Sample 4).

In Fig. 1(b), we show as an example a $\theta$-$2\theta$ XRD scans for Sample 2 synthesized on (111) YSZ. The intense narrow peaks come from the substrate, while the other reflections come from the film and demonstrate that the film is single phase. Only 00*l* reflections with even *l* are observed indicating the (001) orientation of the film and consistent with the *P*6$_3$*cm* space group as shown in Fig. 1(a). Similar patterns are observed for the remaining samples [14]. From STEM images along the [110] zone axis of *h*-LuFeO$_3$ shown in Fig. 1(c), the interface between the film and substrate is seen to be abrupt and free of impurity phases. This film is also found to be nearly free of extra Fe-O layers (syntactic intergrowths of LuFe$_2$O$_4$) [13], which are occasionally observed in similar films [14].

The lattice parameters for each sample were obtained from neutron diffraction measurements of the 300 and 004 nuclear peaks at 5 K as shown in Table 1. These values appear to be independent of both sample thickness and substrate and are within error of previously reported values for stoichiometric films [6,13]. These nuclear peaks are resolution limited [14] indicating complete relaxation of the films with no detectable distribution of lattice parameters in the film or broadening due to finite-size effects. Based on these results it does not appear that the strain potentially induced by the substrate-film interface plays any significant role in determining the overall crystallographic, ferroelectric, or magnetic properties of this system.

Raman measurements, shown for Sample 2 at 10 K in Fig. 2, reveal that the ferroelectric transition occurs well above room temperature. The observed peak positions in the Raman spectra and relative intensities are very similar to that of hexagonal LuMnO$_3$ [15] as opposed to those reported for bulk orthorhombic LuFeO$_3$ [16]; the spectra correspond to that of *h*-LuFeO$_3$. We are able to distinguish at least 10 phonon modes out of the 23 that are active in the scattering geometries used, consistent with its ferroelectric structure [14]. The temperature dependence of the integrated Raman intensity of the strongest $A_1$ peak, near 655 cm$^{-1}$ [normalized by the Bose factor $n + 1 = 1/(1 - e^{-\hbar\omega/kT})$] is shown in the inset. The intensity decreases linearly with increasing temperature between 400 K and 1000 K, above which no change is observed; a linear fit of the intensity over this temperature region demonstrates a clear transition to a non-polar phase at $T_c$ = 1020 K ± 50 K. Piezoelectric force microscopy measurements have shown switching of the ferroelectric polarization in these films [14], consistent with ferroelectricity above room temperature similar to the hexagonal manganites, and confirms previous reports of such behavior in this material [6, 8].

Bulk magnetization measurements indicate that the onset of ferromagnetic order is not coincident with the ferroelectric transition, as expected, and occurs well below room temperature. Measurements of the magnetization along the *c*-axis of Sample 2 are shown in Fig. 1(d). A clear transition can be seen in the field-cooled data at 143 K. While the magnitude of the magnetization is small (~ 0.02 $\mu_B$), it is clear evidence for weak ferromagnetism occurring in the ordered phase. The presence of weak ferromagnetism is common among all samples, with onset temperatures between 140 K and 146 K [14]. The offset between FC and ZFC at higher temperatures in some samples is indicative of trace amounts of Fe$_3$O$_4$ or similar impurity phase which occurs in conjunction with syntactic intergrowths seen in STEM [14, 14]. Beyond this, we find no evidence for additional magnetic transitions at or above room-temperature in measurements of the magnetic susceptibility, shown in the inset of Fig. 1(d). Similar susceptibilities have been measured for magnetic fields applied parallel to the plane of the film indicating isotropic magnetic susceptibility above room temperature, in contrast to a previous report that suggested magnetic order occurs at 440 K [6].

We now turn our focus to the magnetic neutron diffraction results, which provide a full picture of the corresponding antiferromagnetic order. We measured the temperature dependence of several reflections including the 101, 100, and 102, which are predominantly (or entirely) of magnetic origin. The magnetic scattering for all samples is not observed at any temperature above 155 K, consistent with the magnetometry results. Interpretation of the differences among the temperature dependence of these reflections, however, first requires an understanding of the possible **q**=0 magnetic structures consistent with the $P6_3cm$ space group.

When materials order magnetically through second order phase transitions, the types of magnetic structures possible are constrained by the underlying symmetry of the crystal lattice [17]. For the isostructural hexagonal-manganites, representational analysis reveals that the magnetic and crystallographic unit cells of these materials are identical (**q**=0) and that the magnetic structure of these materials may fall into six representations: four one-dimensional and two two-dimensional [18,19]. For the hexagonal manganites, the materials measured thus far have been well described by the one-dimensional representations which contain the classic 120° arrangement of spins in a given plane, labeled as $\Gamma_1$ to $\Gamma_4$ as shown in Figs. 3(a)-3(d), respectively. Planes can either be coupled ferromagnetically ($\Gamma_3$ and $\Gamma_4$) or antiferromagnetically ($\Gamma_1$ and $\Gamma_2$) along the *c*-axis, and the moments may lie along the *a* or *b* crystallographic axes with respect to the 120° arrangement of the spins. Furthermore, only the $\Gamma_2$ representation allows for a net moment to develop along the *c*-axis. Unfortunately, the Fe atoms lie at the $\left(\frac{1}{3}00\right)$ position, such that the $\Gamma_1$ and $\Gamma_3$ representations form a homometric pair, as do the $\Gamma_2$ and $\Gamma_4$ representations. Members of the same homometric pair cannot be distinguished by unpolarized neutron scattering [18].

Different homometries may still be distinguished through unpolarized diffraction by the presence of the 100 magnetic reflection which is found only for the $\Gamma_1$ and $\Gamma_3$ representations.

In Fig. 4 we show detailed neutron diffraction results for Samples 2 and 3 as examples, noting that similar measurements were made on all four samples [14]. In Figs. 4(a) and 4(d), we show that the 101 reflection is present at low temperatures for films grown on both YSZ and $Al_2O_3$ substrates, and clearly absent above the transition temperature determined by magnetometry, again well below that previously reported [6]. Our measurements show no evidence of magnetic scattering above room temperature [14]. The temperature dependence of the scattering intensity of several magnetic reflections measured on warming is shown in Figs. 4(c) and 4(f). The antiferromagnetic ordering temperature $T_N$ is determined by fitting the 101 and 102 reflections with a mean-field order parameter, from which we find $T_N$ = 155 K ± 5 K (Sample 1), 149.7 K ± 1 K (Sample 2), 140.3 K ± 2 K (Sample 3), and 139.6 K ± 1 K (Sample 4) [14]. These values agree quite well with the onset of ferromagnetism obtained from magnetometry, indicating that both in-plane magnetic order and canted moments develop simultaneously and only well below room temperature in these stoichiometric $h$-LuFeO$_3$ films.

Significant intensity is also observed at the 100 reflection at 5 K for these samples as seen in Figs. 4(b) and 4(e). From the temperature dependence of this reflection, however, it is apparent that this does not appear simultaneously with the 101 and 102 magnetic reflections. Rather, the appearance of the 100 reflection is consistent with a reorientation of the moments within the $hk0$ plane below $T_N$, which has been observed in similar systems [19]. The reorientation temperature, $T_R$, is again determined from a fit of a mean-field order parameter of the 100 scattering intensity, from which we find $T_R$ = 53 ± 3 K (Sample 2) and 38 ± 3 (Sample 3), while no such reorientation is discernable for Sample 1 or Sample 4 [14]. For Samples 2 and 3, the ground state magnetic structure can be described by a combination of the $\Gamma_1 + \Gamma_2$ representations as shown in Fig. 3(e), and is consistent with that suggested based on previous measurements [6], while a single $\Gamma_2$ representation alone is adequate to describe Sample 4.

As no spin reorientation was observed in Sample 4, we may more easily refine the magnetic structure including the magnitude of the ordered Fe moments at 5 K from the integrated intensities of several magnetic and structural peaks. These intensities have been corrected to account for the resolution function and appropriately scaled by the intensities of the 004 and 300 nuclear peaks to obtain the proper structure factors listed in Table 2. The magnetic structure is refined using the $\Gamma_2$ representation with the out-of-plane component of the moments fixed to zero, as these magnetic reflections are insensitive to the canted moment. The refinement is in excellent agreement with the data, from which we extract an ordered magnetic moment of 2.9(5) $\mu_B$/Fe. The moment is reduced from that expected for the $S$ = 5/2 $Fe^{3+}$, but follows similar observations of reduced moments in hexagonal manganites [17, 18]. While this refinement was performed using only a single magnetic domain, it should be noted that including equal populations of magnetic domains as discussed in Ref. [12] also resulted in an adequate refinement of the data, and with a comparable magnetic moment of the Fe site.

We conclude that films of metastable $h$-LuFeO$_3$ can be stabilized on different substrates. They exhibit robust magnetic order at a temperature that is substantially below room temperature and below the value (440 K) previously reported [6]. Furthermore, the high temperature magnetic structure in the ordered state of *h*-LuFeO$_3$ does not depend strongly on the underlying substrate; the differences (in $T_N$, for example) that do occur are more likely attributable to compositional differences rather than to the effects of strain. On the other hand, observations of greater variation in the spin-reorientation at $T_R$ (well below $T_N$) is consistent with a more subtle variation of the crystalline structure between films [20], namely the relative displacement the Fe ions within the O-bipyramid [21]. The universal appearance of a ferromagnetically-canted antiferromagnet in the ferroelectric state indicates that the films contain the proper symmetries in the ordered state to support coupling of the ferromagnetic moment directly to an electric field as theoretically proposed [12]. In future work, it will prove interesting to determine whether this observed canted magnetic moment is indeed switchable with electric field, since the films are ferroelectric at temperatures well above room temperature. If so, then further efforts will be warranted to determine if it is possible to increase the magnetic transition temperature in this system, or whether the lessons we learn from this material can be applied in the hunt for materials with similar magnetic properties, but with higher transition temperatures.

**Acknowledgements**

The authors would like to thank Hena Das and Craig Fennie for helpful discussions. Research supported by the U.S. Department of Energy, Office of Basic Energy Sciences, Division of Materials Sciences and Engineering, under Award No. DE-SC0002334. This work made use of the electron microscopy facility of the Cornell Center for Materials Research with support from the National Science Foundation (NSF) Materials Research Science and Engineering Centers program (DMR 1120296) and NSF IMR-0417392. This work was performed in part at the Cornell NanoScale Facility, a member of the National Nanotechnology Infrastructure Network, which is supported by the National Science Foundation (Grant ECCS-0335765). Raman studies at Boise State University have been supported by NSF under grant DMR-1006136, and by M. J. Murdock Charitable Trust "Partners in Science" program (E. L. T.).

Table 1. Lattice parameters for each sample from measurements of the nuclear 300 and 004 film peaks below 10 K.

| Sample | $a$ (Å) | $c$ (Å) |
|---|---|---|
| 1 (200 nm on YSZ) | 5.989(5) | 11.70(3) |
| 2 (250 nm on YSZ) | 5.979(5) | 11.81(3) |
| 3 (200 nm on $Al_2O_3$) | 5.985(5) | 11.77(2) |
| 4 (250 nm on $Al_2O_3$) | 5.994(5) | 11.78(2) |

Table 2. Refinement of the magnetic structure factors measured from integrated intensities at 5 K for Sample 4 grown on $Al_2O_3$. The statistical agreement is given by a $\chi^2 = 1.6$.

| Reflection | $F^2_{calc}$ | $F^2_{obs}$ |
|---|---|---|
| 100 | 0 | 0.01(2) |
| 101 | 4.9 | 4.7(3) |
| 102 | 2.0 | 2.7(4) |
| 201 | 3.1 | 2.9(4) |

**Figure Captions**

FIG. 1 (Color Online) Characterization of a 250 nm thick film of $h$-LuFeO$_3$ on YSZ (Sample 2). (a) Schematic of the crystal structure of $h$-LuFeO$_3$ with the $P6_3cm$ space group. (b) XRD at room temperature, with $h$-LuFeO$_3$ 00$l$ reflections labeled accordingly; substrate peaks are denoted by (*). (c) STEM image of the film near the interface between $h$-LuFeO$_3$ and the substrate. (d) magnetization under FC (closed circles) and ZFC (open circles) conditions. Inset: the high temperature magnetization for magnetic fields 0.01 T, 0.05 T, and 0.1 T applied parallel to the $c$-axis.

FIG. 2 (Color Online) Raman spectra of 200 nm thick $h$-LuFeO$_3$ (Sample 1) measured at 10 K using both polarizations [14], demonstrating the Raman active phonon modes in the ferroelectric state of $h$-LuFeO$_3$. Inset: Normalized Raman intensity of the $A_1$ mode (peak around 650 cm$^{-1}$) as a function of temperature. The red line is a linear fit over the temperature region 400 K < T < 1050 K.

FIG. 3. (a)-(d) Illustration of the four possible one-dimensional representations for $h$-LuFeO$_3$. The possible magnetic structures below T$_R$ consisting of a combination of representations in (a)-(d) are shown in (e) and (f). Labels in parenthesis refer to the equivalent notation used in Ref. [10], for example.

FIG. 4 (Color Online) Neutron diffraction results for (a-c) 250 nm thick films on YSZ (Sample 2) and (d-f) 200 nm thick films on Al$_2$O$_3$ (Sample 3). The intensity for the magnetic 101 reflection is shown in (a) and (d) for each sample, respectively, while the 100 is shown (b) and (e). Solid lines are Gaussian fits through the data. (c) and (f) shown the temperature dependence of the 100, 101, and 102 reflections; the mean-field order parameter is fit for each reflection and shown as the solid line.

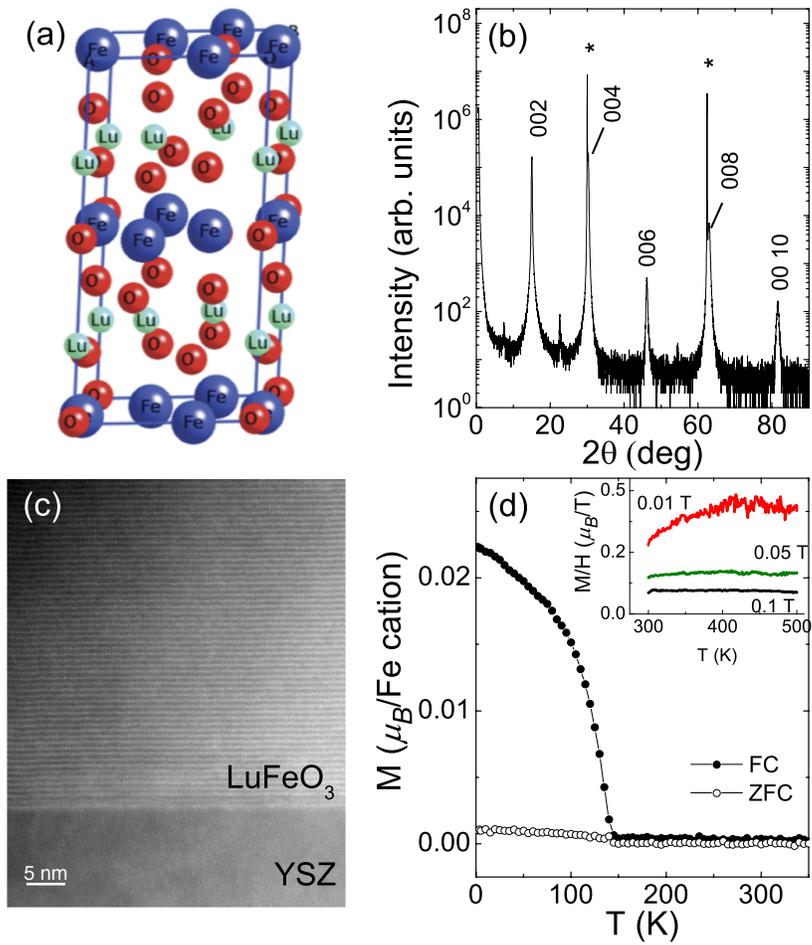

FIG. 1.

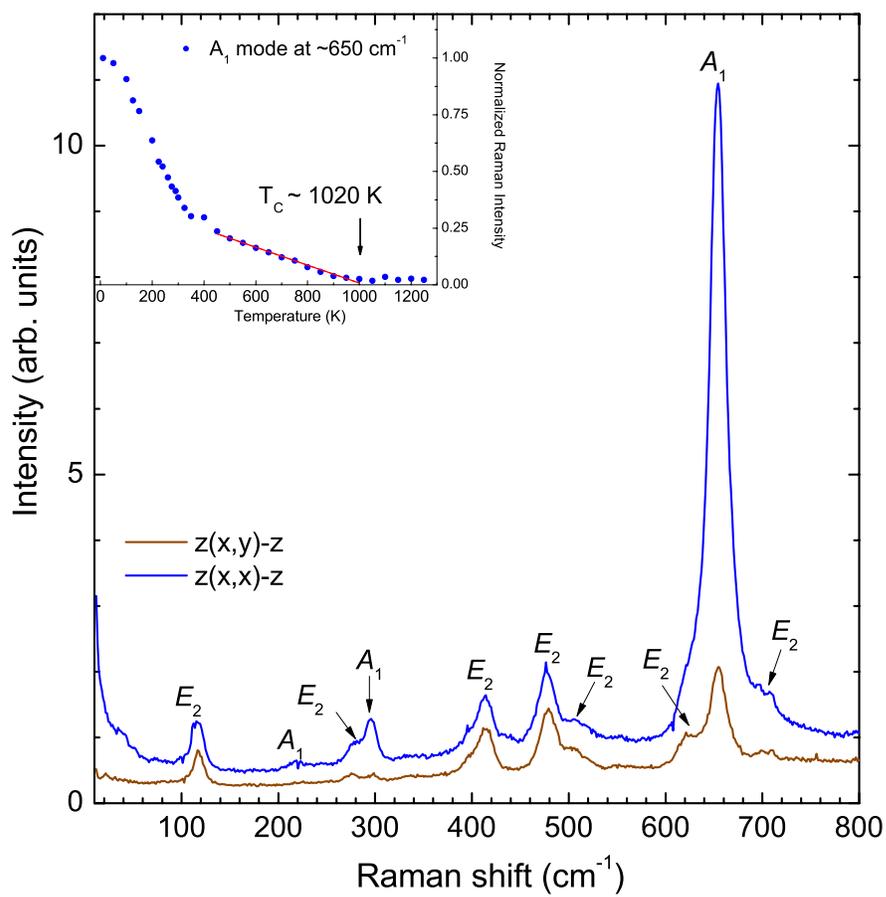

Fig. 2

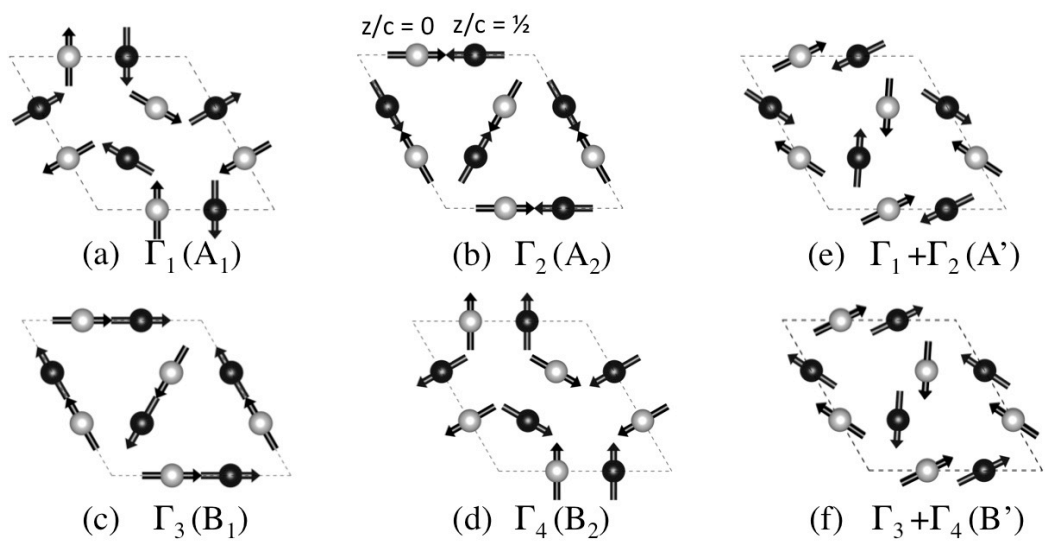

FIG. 3.

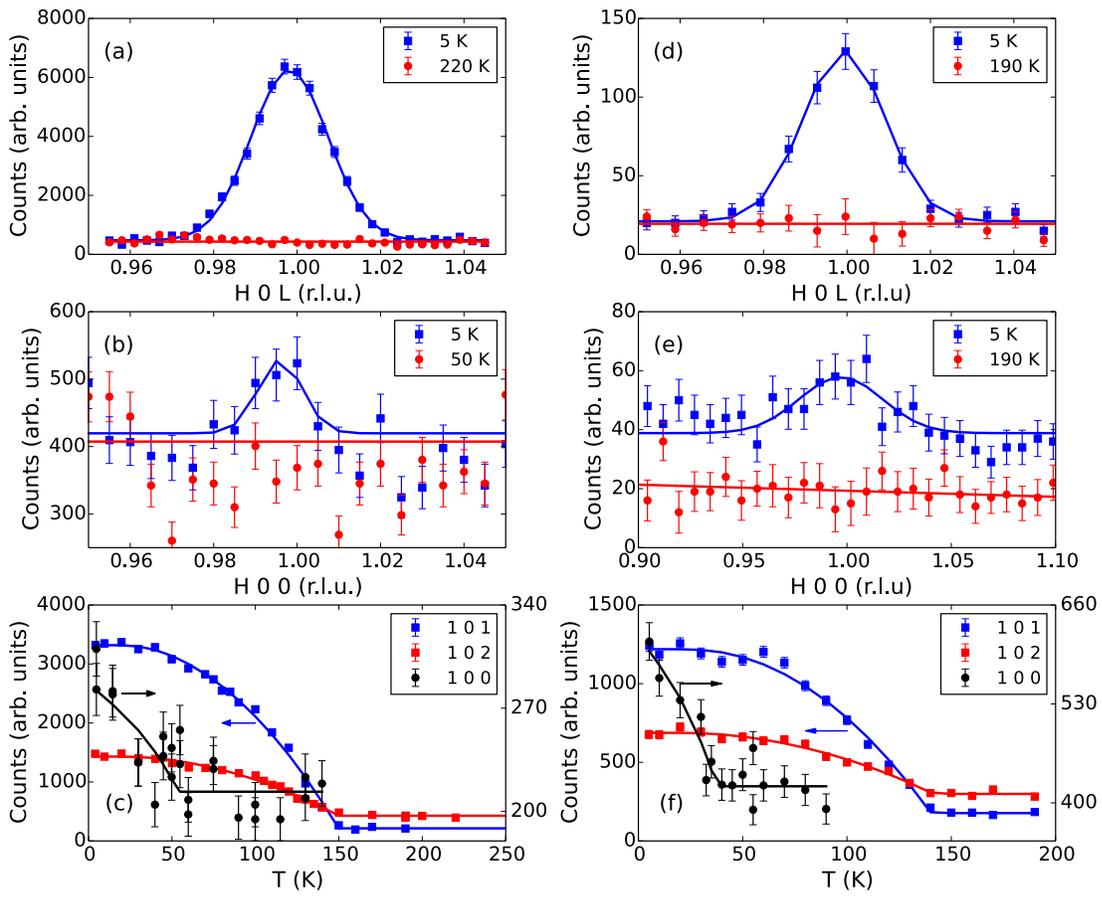

FIG. 4.

## Supplemental Information

**Measurement Details and Characterization of *h*-LuFeO$_3$ Films**

High-resolution X-ray diffraction (XRD) using a four-circle diffractometer along with high-angle annular dark field scanning transmission electron microscopy (HAADF-STEM) were used to characterize the microstructure and assess the quality of the films. The magnetic moment along the *h*-LuFeO$_3$ *c*-axis was measured with a SQUID magnetometer upon warming in a small field (0.01 T) after first cooling in either zero field (ZFC) or in an out-of-plane field of 0.1 T (FC). Raman spectroscopy was performed on a single sample, Sample 1, between 10 K and 1250 K. Neutron diffraction experiments were performed on the BT-4 thermal triple axis at the NIST Center for Neutron Research to determine the magnetic structure and to further characterize the temperature dependence of the magnetization. For neutron diffraction, individual films were measured separately to reduce errors arising from the stacking of many such films together and to reduce the effect of inhomogeneities in films fabricated at different times.

STEM analysis was also performed on a 200 nm film grown on (0001) Al$_2$O$_3$, (Sample 3) as shown in Fig. S1. The substrate-film interface is seen to be sharp, single phase, and and free of double iron oxide layers (syntactic intergrowths of LuFe$_2$O$_4$). A general survey of the entire film thickness indicates the presence of such occasional intergrowths, as seen in the inset of Fig. S1. These are found to occur more frequently in this sample (Sample 3) that in the sample grown on YSZ (Sample 2) shown in Fig. 1.

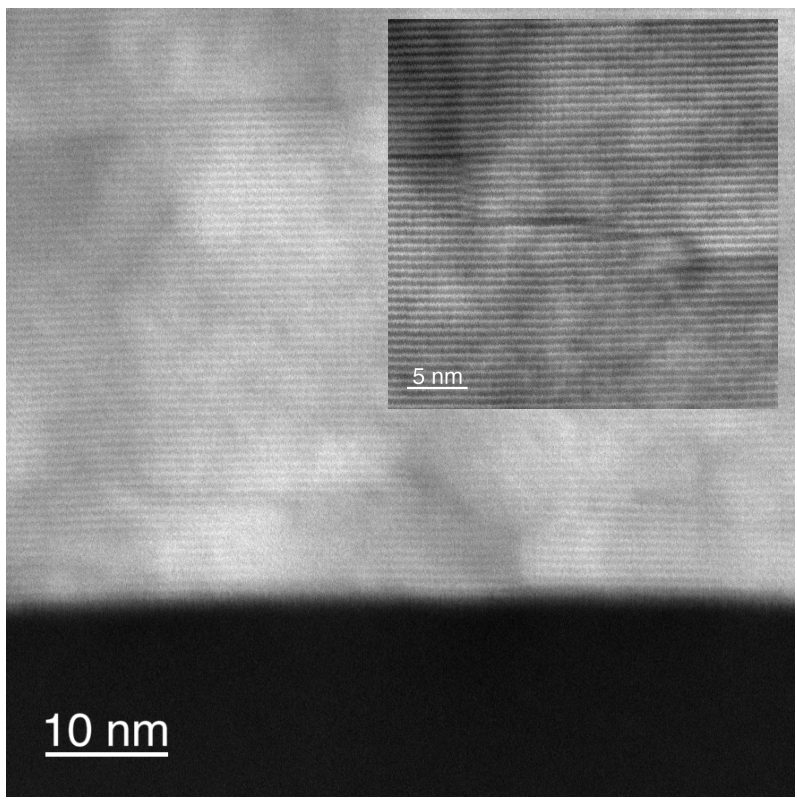

**Fig. S1.** STEM images of Sample 3 on Al$_2$O$_3$.

X-ray diffraction was performed on all films prior to SQUID magnetometry or neutron diffraction measurements. A 220 Ge monochromator was place before the sample to select only Cu $_{K\alpha 1}$ radiation. As with Sample 3, only the even 00$l$ refelctions are observed in the remaining films shown in Fig. S2. Therefore, films are oriented with the $c$-axis normal to the substrate surface regardless of substrate used. The lack of odd 00$l$ reflections is consistent with the $P6_3cm$ structure.

SQUID magnetometry was also performed on Samples 1,3, and 4. The results are shown in Fig. S2. For measurements of samples fabricated on YSZ, a bare YSZ substrate was also measured to allow for the subtraction of paramagnetic and diamagnetic backgrounds. The magnetic signal from the $Al_2O_3$ substrate is sufficiently small that there was no need for this subtraction. From the onset of the ferromagnetic moment, we find transition temperatures of 146 K, 140 K, and 145 K for Samples 1, 3, and 4, respectively. All are above the transition reported from magnetometry in [1] and indicate the bulk of the films are very close to ideal stoichiometry [2].

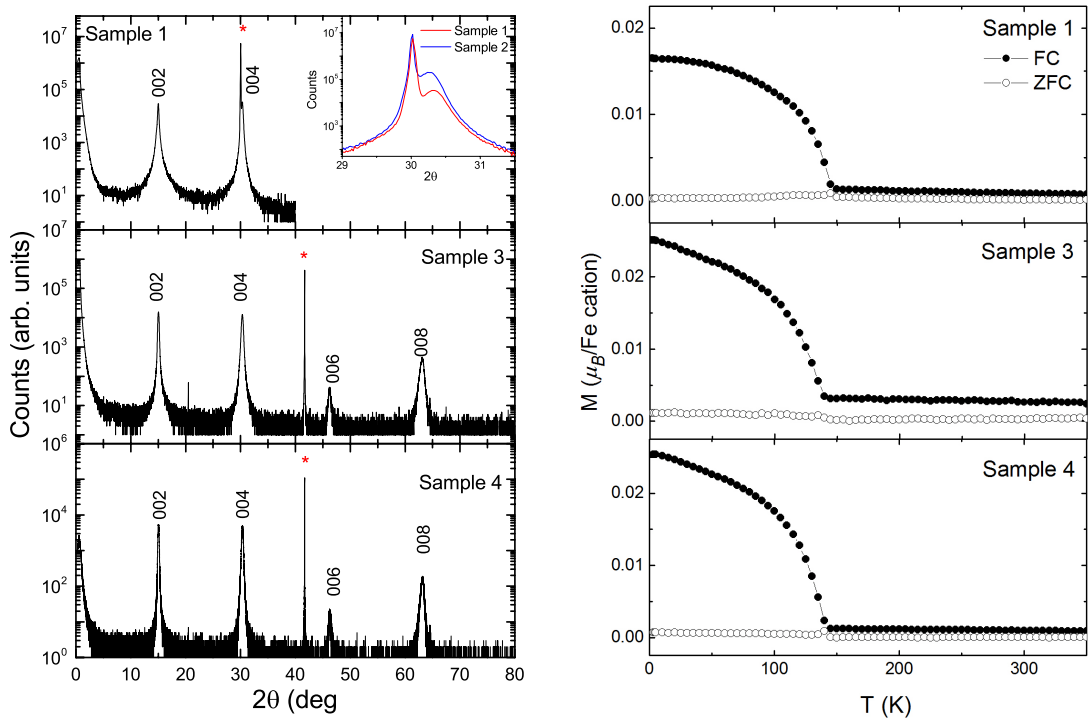

**Fig. S2 (Left)** XRD of Samples 1,3, and 4; 00*l* reflections are labeled accordingly while those arising from the substrate are labeled with an asterisk (*). **(Right)** Magnetic moment along the *c*-axis as a function of temperature for Samples 1,3, and 4. FC data is represented by closed circles and ZFC as open circles.

**Neutron Diffraction and Strain Analysis Parameters**

Pyrolitic Graphite (PG) filters were placed before and after the sample to remove $\frac{\lambda}{2}$ contamination from the beam. Both a monochromator (PG) and analyzer (PG) were employed to ensure elastic scattering (to within instrumental resolution). 40' collimators were employed before and after the monochromator as well as between the sample and the analyzer. Collimations were open between the analyzer and the detector. Order parameter measurements were performed with neutrons of incident energy $E_i$ = 14.7 meV, while magnetic structure determination was performed with $E_i$ = 35 meV. The sample was placed on a single crystal Si wafer to reduce background inside of an aluminum can sealed with He exchange gas to ensure thermal equilibration. Measurements were performed in a closed cycle refrigerator.

We determine the possibility of coherent strain in the lattice by examining the broadening of the nuclear peaks for both $\theta - 2\theta$ as well as rocking curves with fixed $2\theta$ for Sample 4. In Fig. S3(b) and S3(d) we compare the measured $\theta - 2\theta$ intensity with that calculated based the instrument collimations listed above and an ideal sample mosaic $\eta_s$, less than 30', normalized to the measured integrated intensity. These curves are in excellent agreement indicating the measurement is resolution limited such that there is no distribution of the lattice parameters along either the *a* or *c* directions. This suggests that the films are not coherently strained throughout the sample and that strain between the substrate and the *h*-LuFeO$_3$ film is likely contained within a very narrow region. The in-plane sample mosaic, $\eta_s$, can then be approximated from the FWHM of the rocking curves shown in Fig. S3(a) and S3(c). The mosaic is found to have some directional dependence as $\eta_s$ = 82' for the 004 reflection and $\eta_s$, = 67' for the 300 reflection.

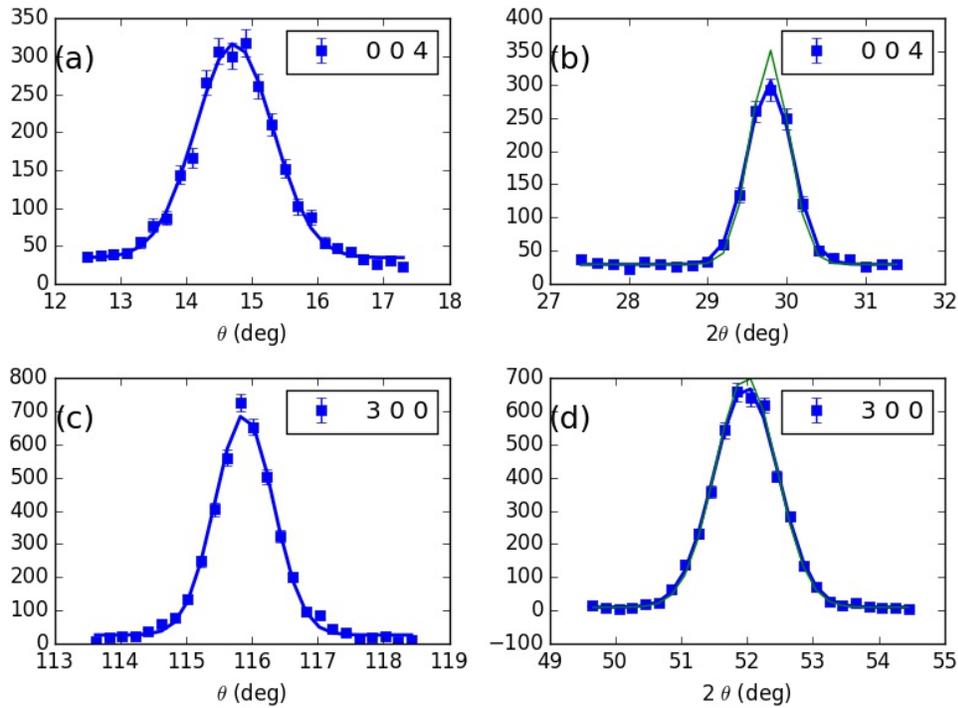

**Fig. S3** (a, c) Rocking (θ-scans) of the 004 and 300 reflections. (b, d) $\theta-2\theta$ scans of the same nuclear reflection, blue curves are fits to the data, while green are calculated based on known instrument resolution.

**Magnetic Order Parameters**

The intensity of the neutron diffraction of the 102 reflection, which is primarily magnetic in nature, was measured also measured for Sample 1 shown in Fig. S4(a) at base and above room temperature. A well defined peak is observed at base temperature, however, no intensity above background is observed at or above 300 K, which conclusively shows that no long-range magnetic order is present in this system at these temperatures. The magnetic ordering and reorientation temperature for Sample 1 was determined by measuring the intensity of the 102 reflection as a function of temperature, shown in Fig. S4(b). The 100 was also measured, however no temperature dependence could be distringuished above background and noise. For Sample 4, no 100 reflection was observed to 5 K, so only the 101 magnetic reflection was measured as shown in Fig. S4(c). The values of $T_N$ were found to be $155 \pm 5$ K and $139 \pm 1.5$ K for Samples 1 and 4, respectively.

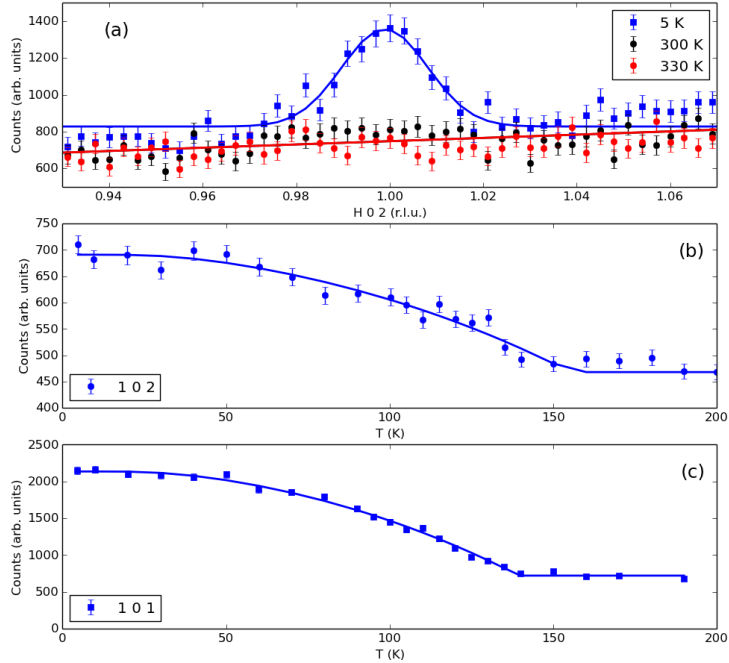

**Fig. S4** (a) Transverse scans through the 102 reflection for Sample 1 both near base temperatures and above room temperature. (b) Temperature dependent intensity of the 102 magnetic reflections for Sample 1, and (c) temperature dependence of the 101 magnetic reflection for Sample 4.

**Raman Scattering**

Raman spectra of a Sample 1, 200-nm thick $h$-LuFeO$_3$ film grown by MBE on a (111) YSZ substrates was measured in backscattering geometry normal to the film surface using a triple spectrometer equipped with a liquid nitrogen cooled multichannel charge coupled device detector. An ultraviolet excitation (325 nm line of He-Cd laser) was used in order to reduce the substrate contribution. The substrate signal was completely suppressed with all signal arising from the LuFeO$_3$ film. Maximum laser power density was 0.5 W/mm$^2$ at the sample surface, low enough to avoid any noticeable local heating of the sample. Spectra for the LuFeO$_3$ sample were recorded in the temperature range 10–1250 K using a variable temperature closed cycle helium cryostat and a high-temperature stage.

Bulk unstrained LuFeO$_3$ has an orthorhombic (distorted perovskite) non-polar *Pbnm* structure (space group $D_{2h}^{16}$) with unit cell containing four formula units. A group theory analysis shows that the sixty phonon modes of orthorhombic LuFeO$_3$ belong to $7A_g + 5B_{1g} + 7B_{2g} + 5B_{3g} + 8A_u + 10B_{1u} + 8B_{2u} + 10B_{3u}$ symmetries [3]. Among these modes, one $A_u$ and two $B_{1u}$ are acoustic. Raman-active are the symmetric modes $7A_g + 5B_{1g} + 7B_{2g} + 5B_{3g}$, of which there are 24 in total.

Hexagonal LuFeO$_3$ was reported to possess a polar *P6$_3$cm* structure with six formula units per unit cell. This structure has 38 Raman active phonon modes ($9A_1 + 14E_1 + 15E_2$), similar to hexagonal manganites of the same symmetry [4, 5]. For the scattering configuration used available in our experiments (backscattering along the *z* direction parallel to the *c*-axis of *h*-LuFeO$_3$), modes of $A_1$ and $E_2$ symmetries are allowed in $z(x,x)\bar{z}$ geometry (parallel polarizations of incident and scattered light), while only $E_2$ modes are active in perpendicular polarization geometry $z(x,y)\bar{z}$.

The temperature evolution of the Raman spectra of the *h*-LuFeO$_3$ film over the range 10–1250 K is shown in Fig. S5. Spectra at elevated temperatures were measured in air in order to avoid decomposition of the films. Therefore, they contain multiple peaks in the low-frequency range (below 200 cm$^{-1}$), which are due to scattering by rotational excitations of molecules of air. The room-temperature spectrum measured after heating up to 1250 K has the same shape and intensity as the spectrum before heating, thus indicating that no noticeable decomposition occurred during heating. The most noticeable change in the Raman spectra with increasing temperature is a decrease of relative intensity of the $A_1$ modes (the peaks seen in Fig. 1 near 296 and 657 cm$^{-1}$). The peak around 655 cm$^{-1}$ is the strongest at low temperatures, its intensity is about 10 times higher compared to the $E_2$ modes at ~415 and 480 cm$^{-1}$. At high temperatures, the $A_1$ peak becomes as weak as those $E_2$ peaks (see Fig. S5(Right) showing enlarged high temperature spectra).

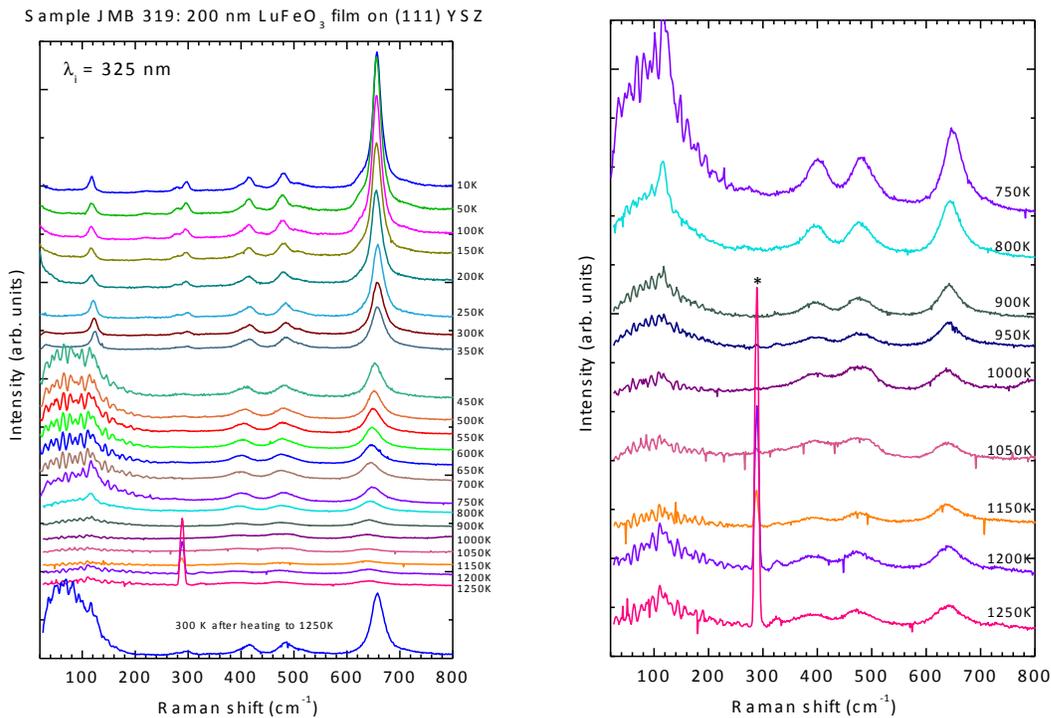

**Fig. S5 (Left)** Raman spectra of a Sample 1 measured in parallel polarization configuration ($z(x,x)\bar{z}$) as a function of temperature. **(Right)** Enlarged part of the left panel showing high temperature spectra. Sharp peak labeled by asterisk (*) is a thermally induced atomic emission line from the copper sample holder.

**Piezoelectric Force Microscopy**

Piezoelectric force microscopy (PFM) was performed to determine the ferroelectric polarizability of the *h*-LuFeO$_3$ in the polar state as determined by Raman spectroscopy. A single-crystal film of platinum was grown on top of YSZ substrate via DC magnetron sputter to act as a bottom electrode for PFM measurements; a 200 nm thick film of *h*-LuFeO$_3$ was then deposited on top of this platinum layer by MBE as previously described. The polarization parallel to the *c*-axis was examined at room temperature by poling the as-grown sample with a ±12 V bias. A 3 μm x 3 μm region was first poled using a -12 V bias such that the polarization points into the plane and shown in Fig. S6 as indicated by the lighter shaded region in the figure. The polarization of a 1 μm x 1 μm box within this poled region was then reversed by applying a voltage of +12 V, indicated as the darker region. This conclusively demonstrates that the material is ferroelectric at room temperature in agreement with Raman spectroscopy and previous measurements, and consistent with previous measurements of thin films fabricated with platinum buffer layers.

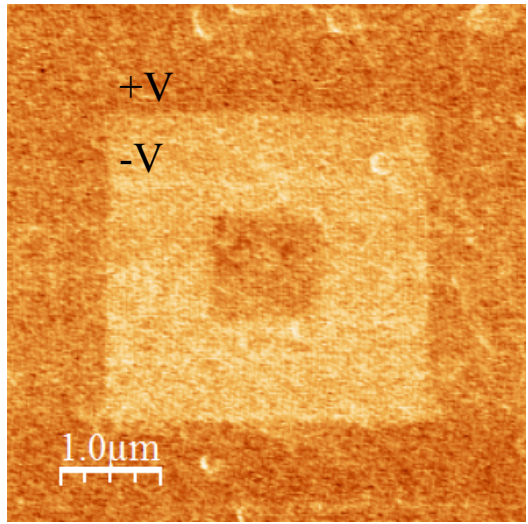

**Fig. S6** An OOP PFM image displaying a box-in-box structure that was written by poling the sample with ± 12V. Darker (lighter) regions correspond to a ferroelectric polarization out of (into) the plane.

**Supplemental Material References**